\documentclass[aps,prl,twocolumn,groupedaddress,showpacs]{revtex4-1}

\usepackage{verbatim}
\usepackage{graphicx}
\usepackage{subfig}
\usepackage{dcolumn}
\usepackage{bm}
\usepackage{xcolor,soul}
\usepackage{amsfonts}
\usepackage{amsmath}

\newcommand{\IM}{\text{Im}}

\newcommand{\qq}{\mathbf{q}}

\newcommand{\kk}{\mathbf{k}}
\newcommand{\rr}{\mathbf{r}}
\newcommand{\w}{\omega}
\newcommand{\bra}{\langle}
\newcommand{\ket}{\rangle}

\begin{document}

\title{Exciton interference in hexagonal boron nitride}

\author{Lorenzo Sponza}
\affiliation{LEM UMR 104, ONERA - CNRS, Ch\^{a}tillon, France}

\author{Hakim Amara}
\affiliation{LEM UMR 104, ONERA - CNRS, Ch\^{a}tillon, France}

\author{Claudio Attaccalite}
\affiliation{CINaM UMR 7325, Aix-Marseille University - CNRS, Marseille, France}

\author{Fran\c{c}ois Ducastelle}
\affiliation{LEM UMR 104, ONERA - CNRS, Ch\^{a}tillon, France}

\author{Annick Loiseau}
\affiliation{LEM UMR 104, ONERA - CNRS, Ch\^{a}tillon, France}

\date{\today}

\begin{abstract}
In this letter we report a thorough analysis of the exciton dispersion in bulk hexagonal boron nitride. 
We solve the \textit{ab initio} GW Bethe-Salpeter equation at finite $\qq\parallel \Gamma K$, which is relevant for spectroscopic measurements.
Simulations reproduce the dispersion and the intensity of recent high-accuracy electron energy loss data. 
We demonstrate that the excitonic peak comes from the interference of two groups of transitions involving the points $K$ and $K'$ of the Brillouin zone.
The number and the amplitude of these transitions determine variations in the peak intensity.
Our results contribute to the understanding of electronic excitations in this system,
unveiling a non-trivial relation between valley physics and excitonic properties.
Furthermore, the methodology introduced in this study to regroup independent-particle transitions is completely general
and can be applied successfully to the investigation of excitonic properties in any system.
\end{abstract}


\maketitle

Hexagonal boron nitride ($h$--BN) is a layered crystal homostructural to graphite.
It displays peculiar optoelectronic properties, measured notably with luminescence~\cite{watanabe-taniguchi-kanda_natmat2004,jaffrennou_jap2007,schue_nanoscale2016,cassabois_natphot2016}, X-rays~\cite{galambosi_prb2011,fugallo_prb2015} or angular resolved electron energy loss spectroscopy (EELS)~\cite{schuster_arxiv2017,fossard_prb2017}.
Several studies have been carried out on its excitonic properties, however some fundamental aspects are still controversial.
For instance, established theoretical calculations predict $h$--BN to be an indirect gap insulator~\cite{arnaud_prl2006,fugallo_prb2015}, and this seems to be confirmed by recent photoluminescence data~\cite{cassabois_natphot2016}, but this conclusion contrasts with the experimental finding of strong luminescence in $h$--BN crystals~\cite{watanabe-taniguchi-kanda_natmat2004}, not compatible with a phonon-assisted excitation picture.
In this context high-accuracy EELS measurements have been performed very recently~\cite{schuster_arxiv2017} at momenta 0.1~\AA$^{-1}$ $\le \qq \le$ 1.1~\AA$^{-1}$ parallel to the $\Gamma K$ direction of the first Brillouin zone.  
The authors give account of an excitonic peak dispersing approximately 0.2~eV, reaching the highest intensity and minimum excitation energy at about 0.7~\AA$^{-1}$ and almost disappearing at 1.1~\AA$^{-1}$.

Finite-momentum EELS give access to the energy and momentum dependent loss function 
\begin{equation}
L(\qq,\w) = \IM[\epsilon(\qq,\w)]/|\epsilon(\qq,\w)|^2 \; ,
\label{eq:loss_function}
\end{equation}
which gives information about the dielectric function $\epsilon(\qq,\w)$ of the probed material.
Peaks of $L(\qq,\w)$ can be put in relation to inter-band excitations ($\propto \IM[\epsilon(\qq,\w)]$) and plasmon resonances ($|\epsilon|\approx 0$). 
So far, measures have been reproduced, interpreted and even anticipated by \emph{ab initio} simulations based on the Bethe-Salpeter equation (BSE) formalism~\cite{strinati}, which includes explicitly the electron-hole interaction (the exciton). 
A very general behaviour, observed in the recent EELS measures~\cite{schuster_arxiv2017} as well, is a sizeable variation of the intensity of $L(\qq,\w)$ as a function of  the exchanged momentum $\qq$, notably the enhancement or the attenuation of excitonic peaks along their dispersion~\cite{gatti-sottile_prb2013,cudazzo_jpcm2015,fugallo_prb2015}.



In this letter we devise accurate numerical methods based on the BSE for the analysis of excitonic features.
We applied them to the investigation of the loss function of bulk $h$--BN in the same energy and momentum conditions as in~\cite{schuster_arxiv2017}, confirming the excitonic nature of the peak, clarifying the origin of its enhancement at 0.7~\AA$^{-1}$ and its dramatic attenuation at higher momentum.
Our analysis provides a deeper insight into the electronic excitations of $h$--BN and it unveils a non-trivial valley physics, indicating possible ways to tune the exciton intensity.
More importantly, the approach introduced here is of general applicability.
We believe that this approach constitutes a helpful way to understand and control excitonic properties in any system.
We are convinced that the outcome of our analysis provides the key ingredients to explain similar effects observed in other materials~\cite{gatti-sottile_prb2013,cudazzo_jpcm2015,fugallo_prb2015}. 
Furthermore, it provides a general methodology to identify how and where the electronic structure has to be modified to achieve the desired exciton intensity.

\subsection{Numerical analysis methods}
EELS and non-resonant inelastic  X-ray scattering give access to the loss function $L(\qq,\w)$ 
with complementary degrees of accuracy in the $\qq$ range~\cite{fossard_prb2017}.

Theory-wise, $L(\qq,\w)$ can be calculated accurately from the dielectric function $\epsilon(\qq,\w)$, obtained as a solution of the BSE.
This can be cast in the form of an eigenvalue problem whose Hamiltonian is most often written in a basis of independent-particle (IP) transitions of index $t=(v,\kk)\to(c,\kk+\qq)$ between occupied and empty states of an underlying IP model, e.g. the Kohn-Sham system. 
Here $(v,\kk)$ indicate the initial state and $(c,\kk+\qq)$ the final state, where $\qq$ is the exchanged momentum laying inside the first Brillouin zone.
Within this framework and including only resonant transitions
\begin{equation}
\epsilon(\qq,\omega) \propto \sum_\lambda \frac{ I_\lambda(\qq)  }{E_\lambda(\qq) - \omega +i\eta}\; ,
\label{eq:epsilon}
\end{equation}
where $E_\lambda(\qq)$ is the energy of the $\lambda$th exciton~\cite{onida-reining-rubio_rmp2002} and $\eta$ a positive infinitesimal quantity.
The spectral intensity
\begin{equation}
I_\lambda(\qq) = \left| \sum_t \tilde{\rho}_t(\qq) A^\lambda_t(\qq) \right|^2= \left|\sum_{t} M_t^\lambda(\qq)\right|^{2}
\label{eq:spectral_weight}
\end{equation}
is the modulus squared of a linear combination of IP-transition matrix elements $\tilde{\rho}_t(\qq)=\bra v\kk | e^{-i\qq\cdot\rr} | c\kk+\qq \ket$ weighted by the exciton wave function components $A^\lambda_t(\qq)$.
The exciton $\lambda$ is called ``bright" when $I_\lambda(\qq)$ is sizeably high, and conversely it is called ``dark" when $I_\lambda(\qq)\approx 0$.
This can happen if either $\tilde{\rho}_t(\qq)$ or $A^\lambda_t(\qq)$ or both are negligible for all $t$, or when IP-transitions interfere destructively leading to a vanishing sum in expression~\eqref{eq:spectral_weight}.
Thus it is sensible to introduce the normalised cumulant weight~\cite{gatti-sottile_prb2013,gogoi-sponza_prb2015}: 
\begin{equation}
\mathcal{J}_\lambda(\qq,E) = \frac{1}{I_\lambda(\qq)} \left|\sum_{t:E_t\le E} M_t^\lambda(\qq) \right|^2
\label{eq:cumulant}
\end{equation}
which allows for a visualization of the building-up of the exciton spectral weight as a function of the IP-transition energy $E$.
This function is positively defined, it tends asymptotically to 1 and in general is not monotonic. 

The normalized cumulant weight~\eqref{eq:cumulant} gives a piece of information relying on the energy of the IP-transitions, though more detailed analysis can be achieved by a careful study of the single $M_t^\lambda(\qq)$ amplitudes themselves.
In particular one can use the phase of $M_t^\lambda(\qq)$ to split IP-transitions into groups depending on their sign in the sum~\eqref{eq:spectral_weight}.
This allows for a deconvolution of the exciton (which includes all IP-transitions) into competing groups of IP-transitions, the intensity of the total peak resulting from the interplay of these contributions.

\subsection{Results}

\begin{figure}
\centering
\includegraphics[width=0.45\textwidth, trim= 25mm 18mm 35mm 18mm, clip]{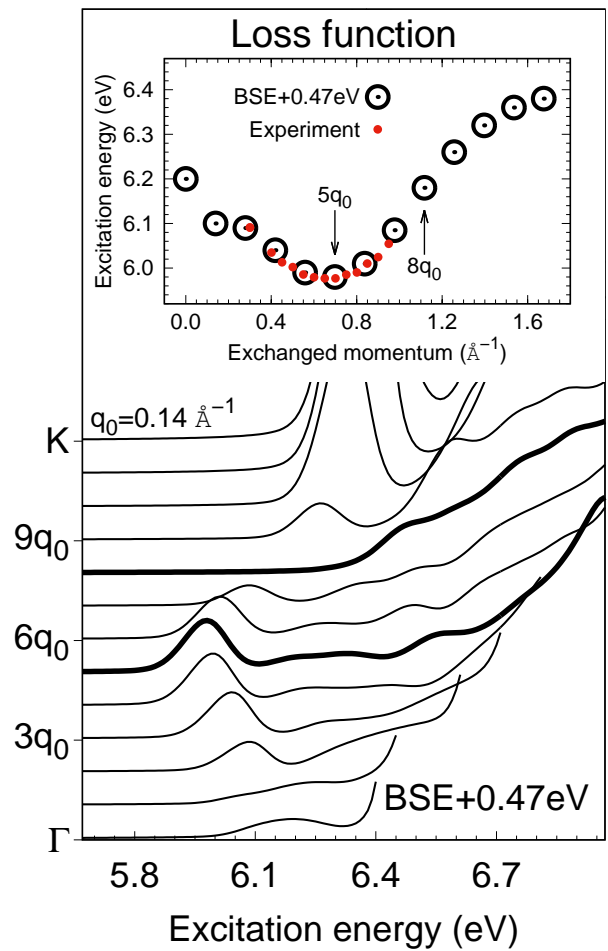}
\caption{(Color online) GW-BSE simulations of the Loss function; thicker lines correspond to $q=5q_0=0.7$\AA$^{-1}$ and $q=8q_0=1.1$\AA$^{-1}$.
Inset: The peak dispersion compared against EELS measurements~\cite{schuster_arxiv2017}.
Simulated data have been blue-shifted by 0.47 eV.}
\label{fig1}
\end{figure}

\begin{figure}
\includegraphics[width=0.45\textwidth,trim= 15mm 15mm 21mm 8mm, clip]{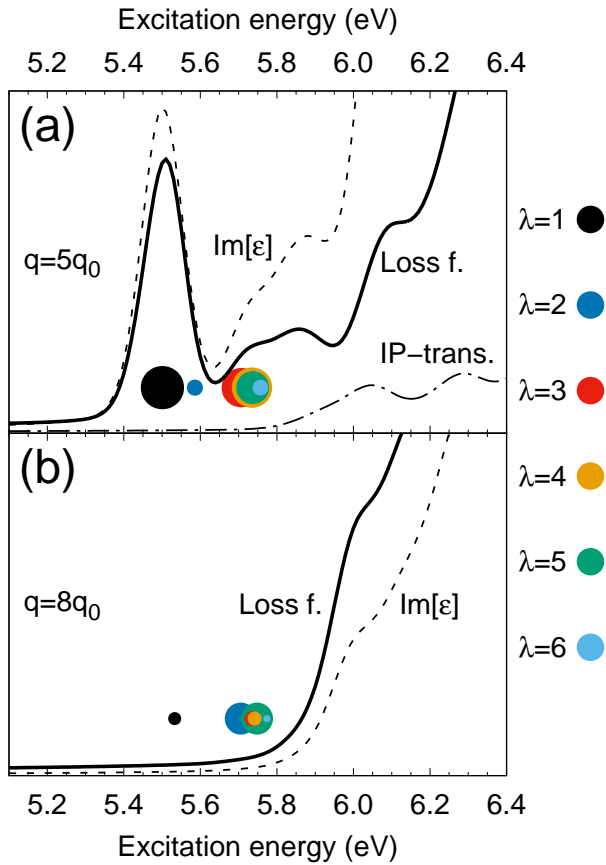}
    \caption{(Color online) (a) Simulated spectra of  $L(\qq,\w)$ (solid) and $\IM[\epsilon]$ with and without electron-hole interaction (dashed and dot-dashed),  at $\qq=5\qq_0\approx 0.7$\AA$^{-1}$. Circles mark the excitonic energies $E_\lambda(\qq)$ for $\lambda \le 6$.
       (b) As in panel (a), but for the exchanged momentum $\qq=8\qq_0\approx 1.1$\AA$^{-1}$. The calculation without electron-hole interaction is not reported.}
\label{fig:eps-dispersion}
\end{figure}

In Figure~\ref{fig1} we report the simulated loss function 
for exchanged momenta $\qq\parallel \Gamma K$ at intervals of $\qq_0=K/12\approx0.14$~\AA$^{-1}$ (see Appendixes). 
In the inset, circles depict the calculated dispersion of the peak compared to 
 the experimental data (red bullets) extracted from~\cite{schuster_arxiv2017}.
Beside a blue-shift of about 0.47~eV that comes from a well-known underestimation of the gap with the G$_0$W$_0$ approximation in this material~\cite{ludger}, the calculated spectra and their dispersion are in very good agreement with the measurements.
In particular our simulations reproduce the fact that the lowest-energy excitation is at $\qq=5\qq_0\approx 0.7$~\AA$^{-1}$, where the peak attains its highest intensity,  and that approximately at $\qq=8\qq_0\approx 1.1$~\AA$^{-1}$ the peak is strongly suppressed (cfr. Figure 1 in~\cite{schuster_arxiv2017}).
At higher $\qq$, the loss function increases again with abrupt intensity, reproducing the strong exciton expected at $\qq=K$ and already analysed elsewhere in literature~\cite{galambosi_prb2011,fugallo_prb2015,fossard_prb2017}.

In this energy range, it turns out that $|\epsilon(\qq,\w)|$ does not vanish, consequently equation \eqref{eq:loss_function} allows us to attribute an interband character to the excitation and put features of the loss function in direct relation to peaks of $\IM[\epsilon]$. 
This appears clearly from Figure~\ref{fig:eps-dispersion}, where we show that $\IM[\epsilon]$ (dashed curve) and $L(\qq,\w)$ (solid curve) at $\qq=5\qq_0$ and $\qq=8\qq_0$ present the same spectral features.
We also mark the energy of the first six excitons, that is $E_\lambda(\qq)$ for $\lambda \le 6$, for both momenta with coloured circles whose size is proportional to $\log[I_\lambda(\qq)]$.
The scale of the loss function in the two panels is the same, and similarly for the scale of $\IM[\epsilon]$. 
Additional information about the dispersion of the first six excitons can be found in Appendix B.

Furthermore, for $\qq=5\qq_0$ we report also the corresponding spectrum of $\IM[\epsilon]$ without electron-hole interaction, i.e. taking into account only independent-particle (IP) transitions between GW levels (dot-dashed curve). 
This spectrum appears flat at 5.5~eV where the BSE calculation predicts a relatively sharp peak.
This comparison confirms the hypothesis, already advanced in~\cite{schuster_arxiv2017}, that the peak has an excitonic nature.

In the following we will focus on the reason of the variations of intensity of the first peak, and in particular at momenta $\qq=5\qq_0$ and $\qq=8\qq_0$, where the intensity is at its highest and its lowest.
Based on the observations done above, we will perform our analysis on $\IM[\epsilon]$ instead of working with the more cumbersome loss function. 


\subsection{Exciton analysis at $\qq=5\qq_0 \approx 0.7$~\AA$^{-1}$}

\begin{figure}[b]
\centering
\includegraphics[width=0.45\textwidth]{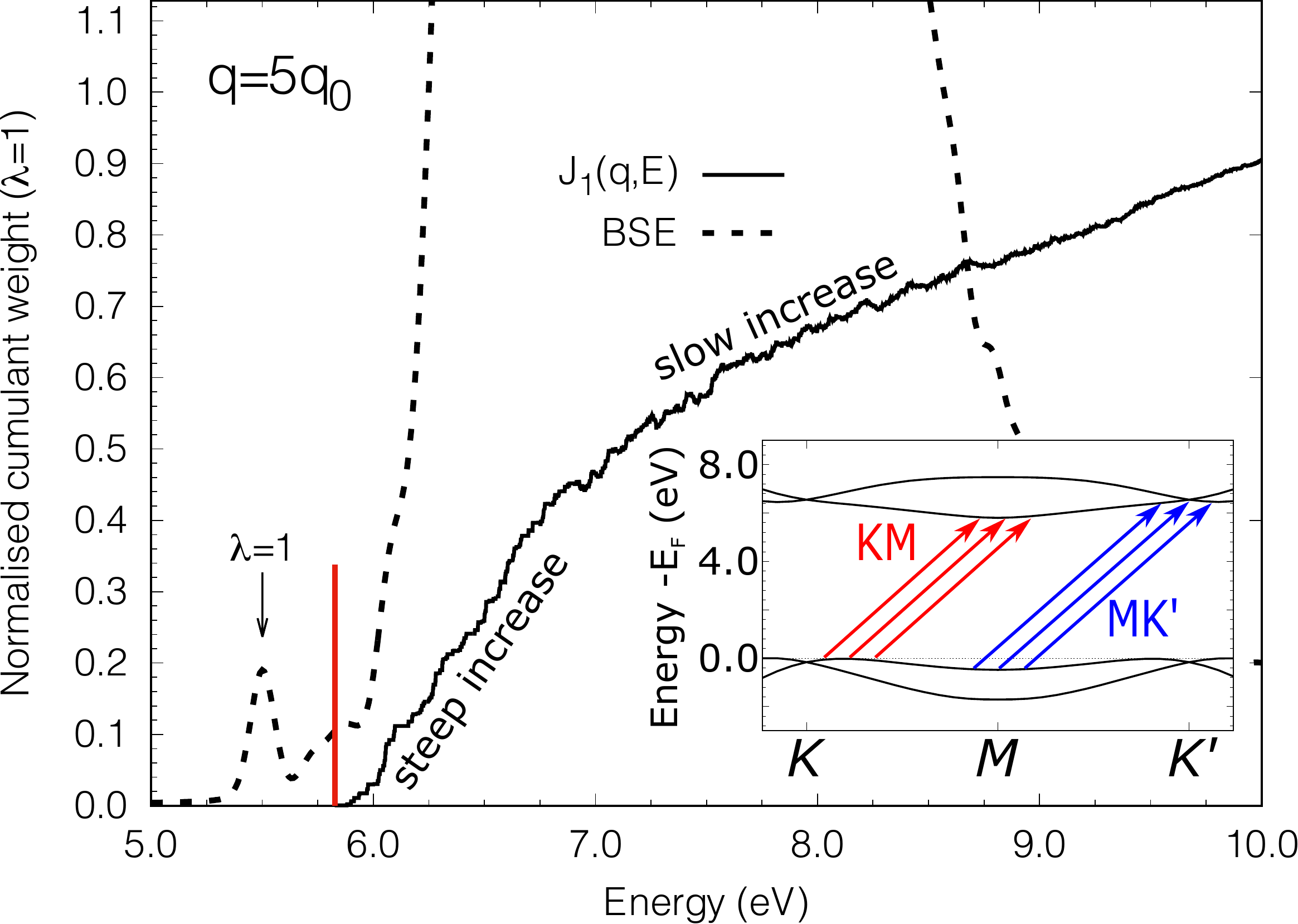}
\caption{(Color online) Cumulant spectral weight (solid curve), GW-BSE spectrum (dashed curve) at $\qq=5\qq_0$. A red vertical line marks the lowest IP-transition.
Inset: The GW band structure along $KMK'$ where the $KM$ and $MK'$ groups of IP-transitions are sketched in red and blue respectively.}
\label{fig3}
\end{figure}

The excitonic peak at $\qq=5\qq_0$ has a binding energy of 0.33~eV, that is the energy difference with respect to the lowest IP-transition with the same $\qq$ (including G$_0$W$_0$ corrections).
In Figure~\ref{fig3} the normalised cumulant weight defined in \eqref{eq:cumulant} is reported versus the energy $E$ of the IP-transitions.
We observe that $\mathcal{J}_{\lambda=1}(\qq,E)$ is a monotonic function of $E$; it rises steeply up to $E\approx 6.8$~eV from where its derivative decreases mildly.
Finally it reaches its asymptotic value of 1 at about $E\approx 12$~eV (not shown).
What this tells us is that IP-transitions sum up constructively at all energies, with most important contributions coming from transitions of energy $E<6.8$~eV.
Indeed these few transitions (0.4\% of the total) account for almost the 42\% of the spectral weight, as $\mathcal{J}_1(5\qq_0,6.8)=0.42$ attests.
Still, to get closer to the full spectral weight, one has to include higher-energy transitions. 
At $E=9.5$~eV, 85\% of the spectral weight is accounted for by still a relatively small number of transitions (less than 10\% of the total).

\begin{figure}
\centering

\includegraphics[height=3.9cm,trim= 24mm 105mm 177mm 32mm,clip]{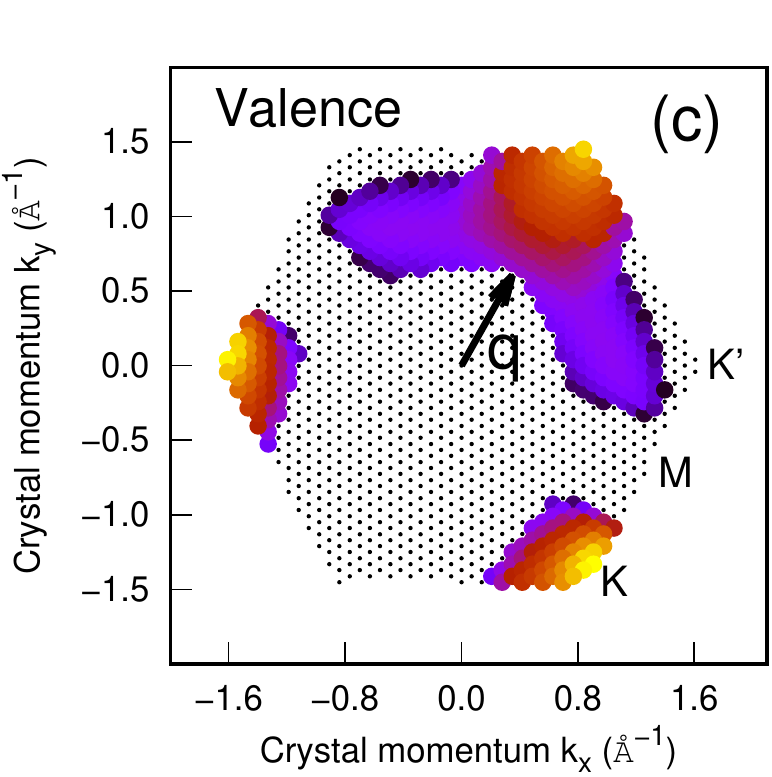}
\includegraphics[height=3.9cm,trim= 160mm 105mm 29mm 32mm,clip]{q5_kp_en000-950_negative}\\

\includegraphics[height=3.9cm,trim= 24mm 105mm 177mm 32mm,clip]{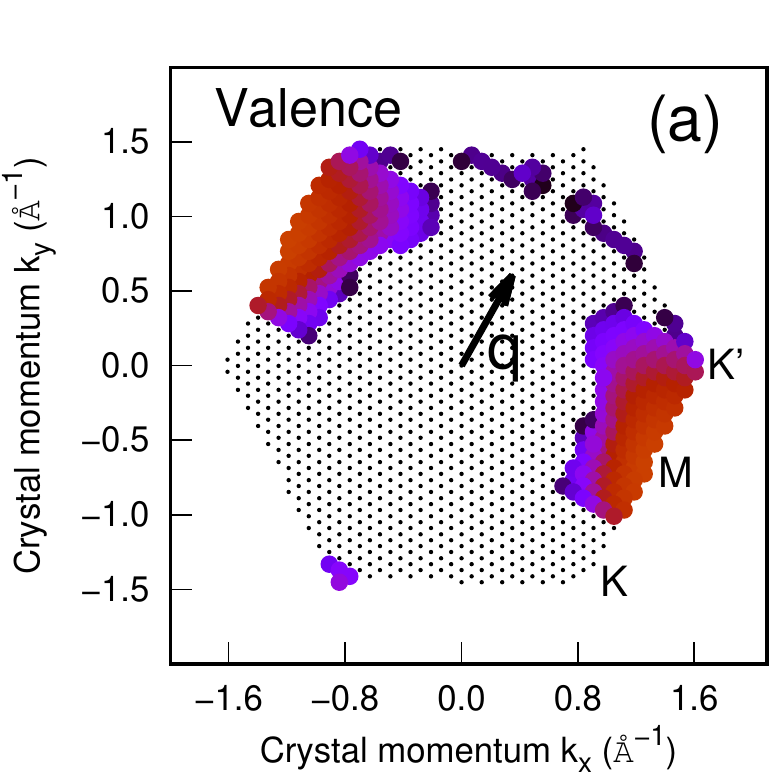}
\includegraphics[height=3.9cm,trim= 160mm 105mm 29mm 32mm,clip]{q5_kp_en000-950_positive}
\caption{(Color online) Top: Transition cartographies of the $KM$ group  at $\lambda=1$ and $\qq=5\qq_0$ plotted for k-points in the $\Gamma K M$-plane for valence (a) and conduction states (b).
Bottom: The same for the $MK'$ group.}
\label{fig:q5K12_analysis}
\end{figure}

We can now gain a deeper insight into the way IP-transitions combine in forming the exciton by looking at the terms of the sum~\eqref{eq:spectral_weight}.
Let us  divide the latter group of transitions ($E\le 9.5$~eV) in three categories: those transitions $t$ for which both real and imaginary parts of the amplitude $M_{t}^\lambda(\qq)$ are positive, those for which both are negative and transitions where they have opposite sign.
The latter group turns out to be composed by transitions with amplitude $M_{t}^\lambda(\qq)\approx 0$, so they do not contribute significantly to the exciton intensity and we can safely neglect them in the analysis.
The other two groups enter the sum of Eq.~\eqref{eq:spectral_weight} with different signs.
Almost one third 
of the considered transitions fall into the first group, with positive amplitude $M^\lambda_{t}(\qq)$.
They are mostly low-energy transitions.
A cartography of these transitions is reported in panels (a) and (b) of Figure~\ref{fig:q5K12_analysis}, where $\log(|\sum_{v,c} M_{(v,\kk)\to (c,\kk+\qq)}^{\lambda=1}(\kk)|)$ is reported as a function of the valence state $\kk$ or conduction state $\kk+\qq$ for k-points in the $\Gamma K M$-plane.
On the other hand, little more than one third of the transitions belong to the negative amplitude group, and they have higher energy but in general lower intensity.
Their maps are reported in panels (c) and (d) of  Figure~\ref{fig:q5K12_analysis}.

The analysis suggests the following interpretation.
Two groups of transitions participate to the formation of the bright exciton ($\lambda=1$), observed in Ref.~\cite{schuster_arxiv2017}.
One group (let us call it $KM$-group) is composed mostly by low-energy transitions going from points close to $K$ to points close to $M$ (and similarly $H\to L$ in the $AHL$-plane, not shown). 
The lowest-energy transitions of this group have also the larger amplitude $M_t^\lambda(\qq)$, and they sum constructively in the steep part of the cumulant ($E<6.8$~eV).
At higher energy, a second group of IP-transitions (call it $MK'$-group), from points in the vicinity of $M$ to points in the vicinity of $K'$ (and $L\to H'$) enter the sum with a negative amplitude, hence canceling partially the contribution of the $KM$ group.
This explains why the derivative of the cumulant decreases from 6.8~eV on, but it is still positive because of the larger number and higher amplitude of the dominating $KM$ group.

The origin of the peak being established, we can now draw the connection with the single-particle band structure.
In the inset of Figure~\ref{fig3} we report the GW band structure along the relevant path $KMK'$, in good agreement with previous calculations~\cite{arnaud_prl2006,galambosi_prb2011,fugallo_prb2015,galvani_prb2016} and experiments~\cite{henck_prb2017}.
The $KM$ and the $MK'$ groups of transitions have been sketched with coloured arrows, respectively red and blue.
At this $\qq$, the $KM$ group of transitions are basically the indirect transitions between the top valence and the bottom of conduction.
The fact that the top valence is close to, but does not coincide with $K$ is consistent with the fact that the lowest excitation is found at $\qq<6\qq_0$.
The strength of the peak is explained by the fact that the $KM$ transitions take place between regions of the band structure where bands are particularly flat (van-Hove singularities). 
Also, the convex curvature of the band structure explains why the $MK'$ transitions start contributing at higher energy and have lower amplitude.

\subsection{Exciton analysis at $\qq=8\qq_0\approx 1.1$~\AA$^{-1}$}

\begin{figure}[b]
\centering
\includegraphics[width=0.45\textwidth, trim= 20mm 20mm 22mm 25mm ,clip]{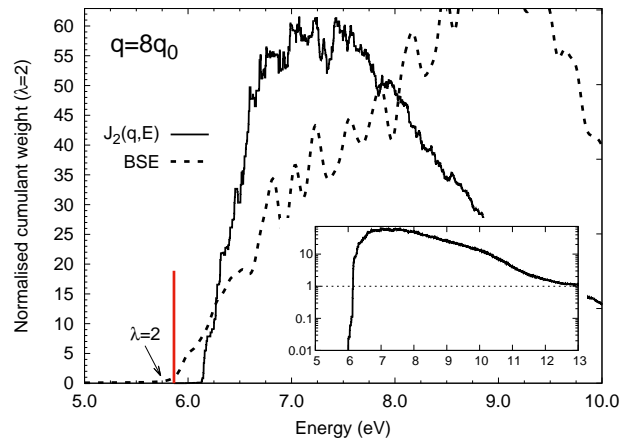}
\caption{(Color online) Cumulant spectral weight (solid curve), GW-BSE spectrum (dashed curve). A red vertical line marks the energy of the lowest IP-transition.
Inset: The cumulant spectral weight up to $E<13$~eV in logarithmic scale.}
\label{fig5}
\end{figure}

\begin{figure}
\centering

\includegraphics[height=3.9cm,trim= 24mm 105mm 177mm 32mm,clip]{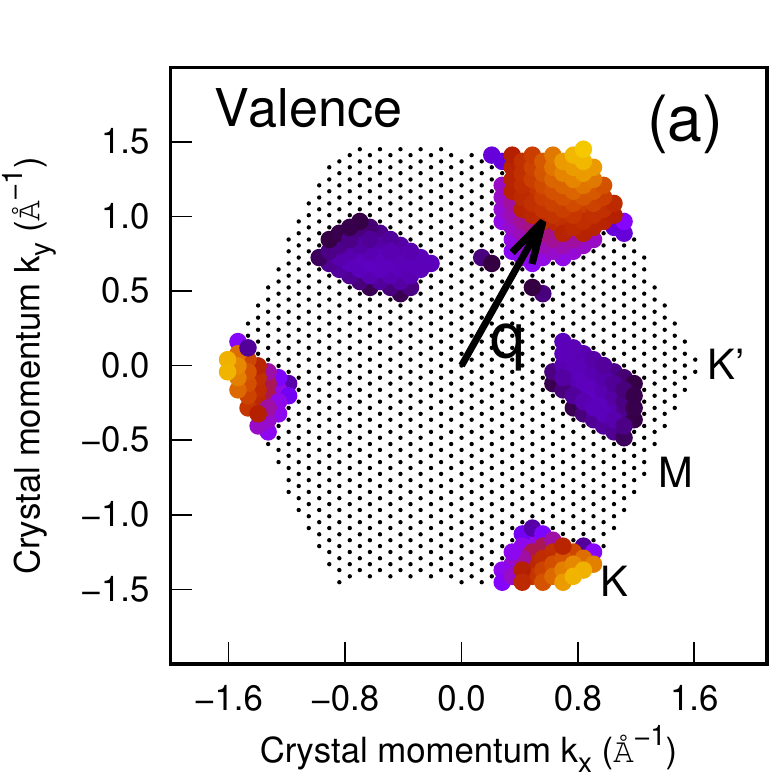}
\includegraphics[height=3.9cm,trim= 160mm 105mm 29mm 32mm,clip]{q8_kp_en000-950_positive}\\

\includegraphics[height=3.9cm,trim= 24mm 105mm 177mm 32mm,clip]{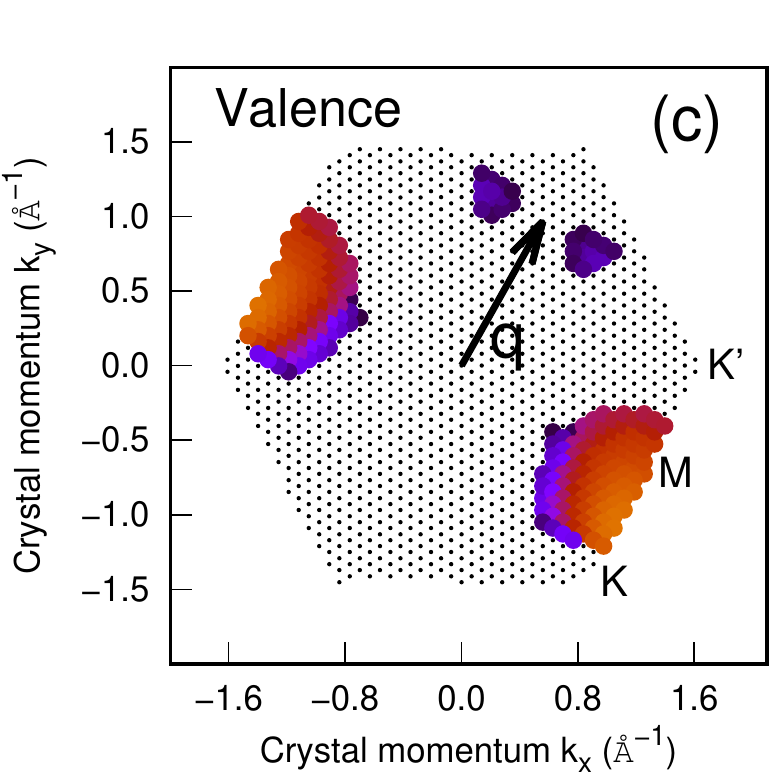}
\includegraphics[height=3.9cm,trim= 160mm 105mm 29mm 32mm,clip]{q8_kp_en000-950_negative.pdf}
\caption{(Color online) Top: Transition cartographies of the $KM$ group at $\lambda=2$ and $\qq=8\qq_0$ plotted for k-points in the $\Gamma K M$-plane for valence (a) and conduction states (b).
Bottom: The same for the $MK'$ group.}
\label{fig:q8K12_analysis}
\end{figure}

Let us switch now to $\qq=8\qq_0$.
At this momentum, the spectral weight is dramatically reduced and it is moved from $\lambda=1$ to a group of higher-energy excitons among which $\lambda=2$ and $\lambda=5$ have the highest (although still very weak) intensity.
In the $\lambda=2$ case, the normalized cumulant weight, reported in Figure~\ref{fig5} does not grow monotonically, instead first it explodes for $E \le 6.8$~eV, where it attains the value of $\sim$ 50, then it attains its maximum between 7 and 7.8~eV and then decreases to reach the asymptotic limit at about 12~eV. 
We can perform the same analysis as before for IP-transitions of energy $E\le 9.5$~eV. 
The cartographies of the IP-transitions are reported in panels (a) and (b) of Figure~\ref{fig:q8K12_analysis} for the $KM$ group, and (c) and (d) for the $MK'$ group.
At variance with the case before the amplitudes of $KM$ and $MK'$ transitions are closer. 

These results can be rationalized as follow.
Most of the IP-transitions entering in $I_{\lambda=2}(\qq)$ up to 6.8~eV are of the $KM$ group and they sum constructively.
But at higher energy, the $MK'$ transitions, which contribute with opposite sign, start having comparable importance.
This induces a halt in the increasing trend ($6.8<E<7.8$~eV) and eventually they dominate bending down the cumulant back to its asymptotic limit.
The result is that the two groups of transitions almost cancel each other, leading to a very weak intensity.


It is worth recalling here that the $\lambda=2$ exciton is almost degenerate with another exciton of non-negligible intensity ($\lambda=5$).
Not surprisingly, carrying out a similar analysis on the latter leads to basically the same results (see Appendix C).

\subsection{Conclusions}
We computed the loss function of  bulk $h$--BN solving the \emph{ab initio} GW-Bethe-Salpeter equation at finite $\qq$ along the $\Gamma K$ direction, which is  relevant for spectroscopic studies.
We observe an excitonic peak dispersing of about 0.45~eV, displaying a strong intensity at $\qq\approx 0.7$~\AA$^{-1}$, where the excitation energy is the lowest, and almost disappearing at $\qq\approx 1.1$~\AA$^{-1}$.
These findings are in very good agreement with recent electron energy loss experiments~\cite{schuster_arxiv2017}.

The associated dielectric function displays similar characteristics.
We show that the peak intensity is determined by the interference of two groups of transitions contributing to the peak formation with opposite signs.
Our investigation allow us to unveil a non-trivial connection between the exciton dispersion, its intensity and the electronic structure in the vicinity of $K$($H$) and $K'$($H'$) points in bulk $h$--BN, eventually suggesting ways to control excitonic properties by changing the electronic structure in the vicinity of the $K$ and $K'$ valleys.
It is worth stressing that with the help of the methodology we devised, it is possible to use spectroscopic methods to probe electronic excitations at the two valleys at the same time. This is of paramount importance, for instance in the vallytronics of layered systems~\cite{yuan_nanolett2016}.

Furthermore, the methodology presented in this work is of general applicability and could be extended to studies of excitonic properties in any system. 
The splitting of relevant IP-transitions into appropriately defined groups can simplify the interpretation of the excitonic properties, help the analysis and possibly disclose some non-trivial mechanism. 
We believe that the strategy adopted here can be employed successfully also to other cases in bulk as well as in 2D materials.
This helps the interpretation of measured data (as the case of our application to $h$--BN) but most importantly it can suggest where and how to change the electronic structure whenever a control on the excitonic intensity is required.

\begin{acknowledgments}
The authors thank Doctor R. Schuster for the clarifications regarding the experimental data \cite{schuster_arxiv2017}.
The research leading to these results has received funding from the European Union H2020 Programme under Grant Agreement No. 696656 GrapheneCore1. We acknowledge funding from the French National Research Agency through Project No. ANR-14-CE08-0018.
\end{acknowledgments}

\medskip

\section{Appendices}

\subsection{Appendix A: Computational details}
The simulated $h$--BN has lattice parameters $a=2.5$~\AA~ and $c/a=2.6$~\cite{solozhenko_ssc1995}.
The Kohn-Sham system and the GW corrections have been computed with the ABINIT simulation package (a plane-wave code~\cite{abinit}). 
Norm-conserving Troullier-Martins pseudopotentials have been used for both atomic species.
DFT energies and wave functions have been obtained within the local density approximation (LDA) to the exchange-correlation potential, using a plane-wave cutoff energy of 30~Ha and sampling the Brillouin zone with a $8\times 8\times 4$ $\Gamma$-centred k-point grid.
The GW quasiparticle corrections have been obtained within the perturbative GW approach. 
They have been computed on all points of a  $6\times 6\times 4$ $\Gamma$-centred grid, a cutoff energy of 30~Ha defines the matrix dimension and the basis of wave function for the exchange part of the self-energy.
The correlation part has been computed including 600 bands and applying the same wave function basis as before.
To model the dielectric function, the contour deformation method has been used, computing the dielectric function up to 60~eV, summing over 600 bands and with a matrix dimension of 6.8~Ha. 
The quasiparticle corrections have been subsequently interpolated on a denser $36\times 36\times 4$ k-point grid where the BSE calculation has been carried out.

The macroscopic dielectric function $\epsilon(\qq,\w)$ has been calculated at the GW-BSE level in the Tamm-Dancoff approximation using the code EXC~\cite{exc}.
We included six valence bands and three conduction bands; 360~eV is the cutoff energy for both the matrix dimension and the wave function basis. 
The static dielectric matrix entering in the BSE kernel has been computed within the random phase approximation with local fields, including 350 bands and with a cutoff energy of 120~eV and 200~eV for the matrix dimension and the wave function basis respectively.
With these parameters, the energy of the first peaks of $\epsilon(\qq,\w)$ are converged within 0.01 eV and their intensity are converged within 5\%.

All reported spectra have been convoluted with a Gaussian of $\sigma=0.05$~eV in order to reduce the noise due to the discrete k-point sampling and to simulate the experimental broadening.

\subsection{Appendix B: Dispersion of the first six excitons}

\begin{figure}[b!]
\centering
\includegraphics[width=0.5\textwidth, trim= 20mm 20mm 20mm 15mm,clip]{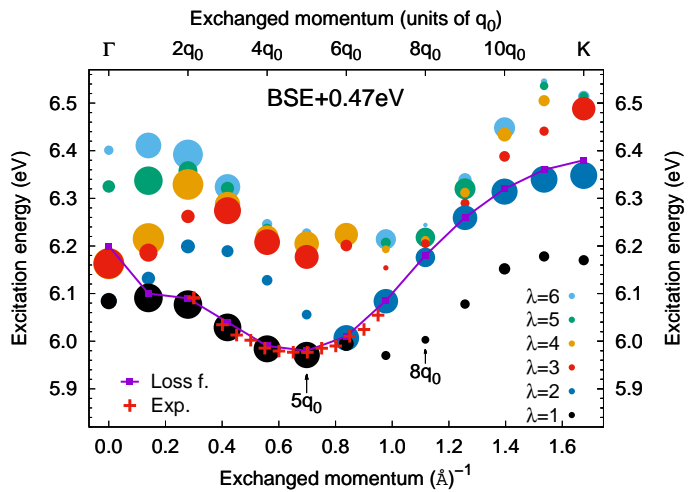}
\caption{(Color online) Dispersion of the excitonic energies $E_\lambda(\qq)$ for $\lambda \le 6$. The size of the circles is proportional to the logarithm of the peak intensity. A Purple line connects the lowest peaks of the loss function $L(\qq,\w)$. Red crosses are the peak of the experimental loss function extracted from~\cite{schuster_arxiv2017}. All calculated points have been blue-shifted by 0.47~eV.}
\label{fig:eps_dispersion_sm}
\end{figure}

In Figure~\ref{fig:eps_dispersion_sm} we report the dispersion of the first six excitons along the line $\Gamma K$ with coloured circles whose size is proportional to $\log[I_\lambda(\qq)]$, so larger circles correspond to bright excitations. 
The points have been obtained within the GW-BSE framework an shifted by 0.47~eV to higher energies.

As expected~\cite{arnaud_prl2006,galvani_prb2016,koskelo_prb2017}, at $\qq=\Gamma$ the first two excitons are degenerate and basically dark, whereas all the peak intensity is concentrated on the degenerate excitons with $\lambda=3$ and $4$ (the two are superimposed in the plot, so that only $\lambda=3$ is visible).
As soon as one moves away from $\Gamma$, the degeneracy is lifted~\cite{koskelo_prb2017} and the first bright peak coincides with the lowest energy exciton ($\lambda=1$).	
This is valid up to $\qq\approx 6\qq_0$ (halfway in the $\Gamma K$ line) where the peak intensity is moved to $\lambda=2$ as a consequence of a band crossing.
The intensity of the excitations is successively reduced at $8\qq_0$ and $9\qq_0$ where several excitons are concentrated in a narrow energy range.
Finally, as $\qq$ approaches $K$, the exciton $\lambda=2$ steps-up again concentrating most of the intensity.

We also report on the same Figure the dispersion of the loss function as measured~\cite{schuster_arxiv2017} (red crosses) and computed in this work (purple squared curve).
One can see that the position of the peak of $L(\qq,\w)$ follows closely the dispersion of the first bright excitation of $\IM[\epsilon]$ (lowest energy larger circles).

\subsection{Appendix C: Analysis of the $\lambda=5$ exciton at $\qq=8\qq_0$} 
\begin{figure}[h]
\centering
\includegraphics[width=0.5\textwidth, trim= 20mm 20mm 20mm 20mm, clip]{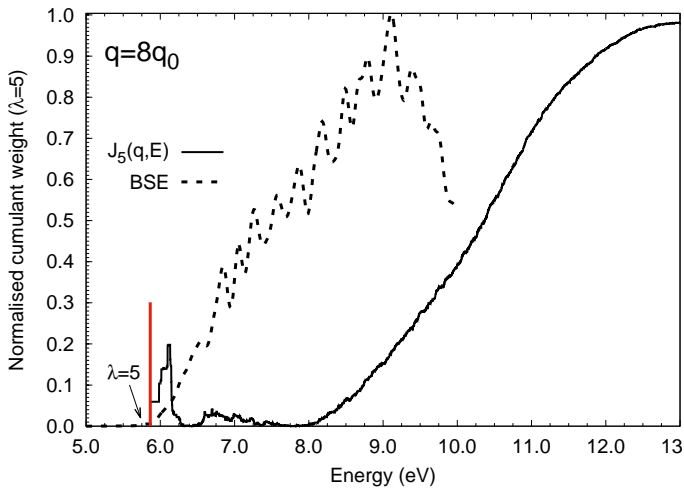}
\caption{(Color online) Cumulant spectral weight (solid curve) of the $\lambda=5$ exciton and the corresponding GW-BSE spectrum (dashed curve) at $\qq=8\qq_0$. A red vertical line marks the energy of the lowest IP-transition.}
\label{fig:cumulant_sm}
\end{figure}

At $\qq=8\qq_0$, the intensity of the peak is very low.
This low intensity is basically shared by two excitons, $\lambda=2$ (analysed in the main text) and $\lambda=5$ with an energy around 50~meV higher.
The analysis with the cumulant weight, reported in Figure~\ref{fig:cumulant_sm}, has a surprising shape.
After a first increase around 6~eV, the cumulant decreases abruptly and vanishes at 6.5~eV. 
Then it oscillates around the value 0 until one starts including IP-transitions of energy $E \ge 8.0$~eV.
Remembering that cumulant weight is defined as a modulus squared, one realizes immediately that what is observed is again a phenomenon of interference (as in the $\lambda=2$ case), but the dominating group changes during the analysis.
At the very beginning ($E \le 6.1$~eV) the $KM$ group constructs the peak, but immediately after the $MK'$ transitions cancel this contribution and leads the cumulant weight back to 0. 
From this point, the two contributions mutually cancel and it is only at $E>8$~eV that the $MK'$ group prevails and the cumulant start growing monotonically to its asymptotic limit.

\end{document}